\documentclass[pra,twocolumn,amsmath,amssymb]{revtex4}
\usepackage{graphicx}
\usepackage{bm}
\usepackage{amsmath,amssymb}
\usepackage{color}
\usepackage{ascmac}

\begin{document}
\title{Testing honesty of quantum server}  
\author{Tomoyuki Morimae}
\email{morimae@gunma-u.ac.jp}
\affiliation{ASRLD Unit, Gunma University,
1-5-1 Tenjin-cho Kiryu-shi Gunma-ken, 376-0052, Japan}

\date{\today}
            
\begin{abstract}
Alice, who does not have any sophisticated quantum technology,
delegates her quantum computing to Bob, who has a 
fully-fledged quantum computer. Can she check whether the computation 
Bob performs for her is correct? 
She cannot recalculate the result by herself, since she does not
have any quantum computer.
A recent experiment with photonic qubits~\cite{BarzNP} 
suggests she can. 
Here, I explain the basic idea of the result,
and recent developments about secure cloud quantum
computing.
\end{abstract}

\maketitle  
First-generation quantum computers will be implemented in the cloud style, 
since only limited groups, such as governments and large companies, 
will be able to possess such expensive and high-maintenance machines. 
How can client's privacy be protected? 
How can the client be convinced of the correctness of the result
although she does not have any quantum computer? 

Imagine that you do online shopping. Of course, you do not want to reveal your 
private information, such as what you bought, your credit card number, 
and your home address, etc. to someone else. Alternatively, imagine that a 
pharmaceutical company uses a time-sharing service of a super-computer 
provided by an electric company to run their own molecular dynamics program. 
The pharmaceutical company must want to make sure that the program, which is 
their top secret, will not be divulged. In short, securing client's privacy 
in cloud computing is one of the most central problems in today's digital 
society. In fact, there has been a long research history in classical 
cryptography on this problem, and several important results 
have been obtained 
(such as the famous fully homomorphic encryption by Gentry~\cite{Gentry}).

In quantum computing, Broadbent, Fitzsimons, and Kashefi~\cite{BFK} 
proposed in 2009 a protocol of secure cloud quantum computing which uses 
the measurement-based quantum computation~\cite{MBQC}. In the protocol, 
Alice needs only a device that emits randomly-rotated
single-qubit states. Bob has a sufficient technology to
conduct the universal measurement-based quantum computing.
Alice sends many randomly-rotated single-qubit states to Bob,
and after that Alice and Bob perform some two-way classical
communications. It was shown in Ref.~\cite{BFK} that
if Bob is honest, Alice can obtain the correct output (correctness),
and whatever the malicious Bob does, he cannot learn anything 
about Alice's input, the program, or the result of the computation
(blindness). 
The protocol was experimentally 
demonstrated with photonic qubits in 2012~\cite{Barz}. 

Plenty of theoretical developments have also been
done since then~\cite{FK,Vedran,AKLTblind,topoblind,
CVblind,topoveri,MABQC,Sueki,composable,composableMA,
distillation,Lorenzo,Joe_intern}.
For example, it was pointed out in Ref.~\cite{Vedran}
that in stead of single-photon states,
Alice has only to generate weak coherent pulses.
It was also shown in Ref.~\cite{MABQC,topoveri}
that single-qubit measurements are also enough for Alice.
The original blind protocol that uses the cluster state
has been generalized to other resource states,
such as Affleck-Kennedy-Lieb-Tasaki (AKLT) state~\cite{AKLTblind},
Raussendorf-Harrington-Goyal topological state~\cite{topoblind},
and the continuous-variable cluster state~\cite{CVblind}, etc.
The communication complexity between Alice and Bob
was also studied~\cite{Lorenzo,Joe_intern}.

\begin{figure}[htbp]
\begin{center}
\includegraphics[width=0.3\textwidth]{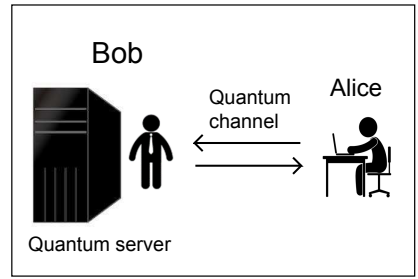}
\end{center}
\caption{
The cloud quantum computing.
Alice delegates her quantum computation to Bob.
} 
\end{figure}

In this way, malicious Bob cannot learn Alice's information. 
However, there is another problem:
he can still tamper with it! If malicious Bob wants to 
fool Alice, he can forge the result, or just deviate 
from the correct procedure. Then, the result of the computation is no 
longer correct, but Alice has to accept the wrong result, since she 
cannot check the correctness of the result by herself (remember that she 
does not have any quantum computer). 
How can Alice avoid such an unpleasant situation? 
In the above example of the pharmaceutical company, 
such a question is crucial, since the pharmaceutical company, who pays a 
huge amount of money for the service, does not want to be palmed off with 
a wrong result by a fishy company trying to sell a fake quantum computer.

A protocol is called verifiable if the probability that the client accepts 
a wrong result can be sufficiently small. 
In Ref.~\cite{FK}, Fitzsimons and Kashefi 
improved the Broadbent-Fitzsimons-Kashefi protocol~\cite{BFK} so that it 
is verifiable. (See also Refs.~\cite{topoveri,Aharonov} for
other verification protocols.)
The basic idea of these verifiable protocols is to hide 
``trap" qubits in the computation.
Bob does not know the position of the traps, 
and therefore he touches a trap and changes its state with a certain 
probability if he deviates from the correct procedure. Alice
checks the trap qubits, and accepts the result of the computation 
only if no trap is altered. The probability that Alice 
accepts a wrong result is the probability that the lucky Bob 
can alter computational information without touching any trap. 
If Alice uses a quantum error correcting code, she
can make such a probability exponentially small: if the computation 
is encoded with a quantum error correcting code, Bob has to apply 
some global operations to alter the logical state, and such a requirement 
of global operations drastically increases the probability of the server 
touching a trap. (An analogy is that
a tank will more likely hit a land mine than a pedestrian,
because the tank sweeps larger space.) 

A full implementation of verification protocols~\cite{FK,topoveri,Aharonov} 
is, however, very challenging with current technology. 
In~\cite{BarzNP}, the authors 
provide a simplified protocol feasible with current technology, and 
demonstrate it using four photonic qubits. The simplification trades off 
some advantages of the original protocol~\cite{FK} (and Ref.~\cite{topoveri}), 
such as the above mentioned exponentially small probability of accepting a 
wrong result, and therefore many runs of protocols are necessary. 
Furthermore, generalizations of their simplified protocol tuned for four 
qubits to non-scalable but many-qubit quantum computers are not clear. 
However, the essential idea that hiding
traps can detect malicious Bob 
is cleverly achieved in spite of the four qubits limitation, 
and therefore their result is the first important proof-of-principle 
demonstration of testing quantum server.

Interestingly, verification is an important concept not only in 
cryptography but also in foundations of physics and computer 
science~\cite{Reichardt,Aharonov}. 
Ultimately, physics is the activity of theoretically predicting a phenomenon, 
and experimentally confirming it. However, the behavior of a quantum 
many-body system is notoriously
too complicated to be efficiently simulated on a classical 
computer, and therefore such a prediction-and-confirmation paradigm will 
break down in the many-body limit. A verification method is expected to 
be a solution: it enables an experimentalist who has only 
limited quantum technology to ``verify" that a quantum many-body system 
in front of her is correctly evolving according to her theory. 

Isn't it very exciting that a study of cryptography 
will shed new light on the foundation of physics?


\end{document}